\documentstyle[aaspp4]{article}

\begin{document}

\title{ENERGY CRISIS IN ASTROPHYSICS \\
(Black Holes vs. N-Body Metrics)}

\author{Carroll O. Alley, Darryl Leiter\altaffilmark{1}, Yutaka
Mizobuchi\altaffilmark{2} and
H\"{u}seyin Y{\i}lmaz\altaffilmark{3}}
\affil{Department of Physics, University of Maryland, \\
College Park, Maryland \ 20742}

\altaffiltext{1}{9 Marwood Drive, Palmyra, Virginia 22963}
\altaffiltext{2}{Central Research Laboratory, Hamamatsu Photonics, K.K., Hamakita, 434-8601 Japan}
\altaffiltext{3}{Hamamatsu Photonics, K.K., Hamamatsu City, 430-8587 Japan and
Electro-Optics
Technology Center, Tufts University, Medford, Massachusetts}

\begin{abstract}
The recent observation of the gamma ray burster GRB 990123, requiring at least two $M_{\sun}
c^2$ of energy in gamma radiation alone, created an energy crisis in astrophysics(\cite{3}).  We
discuss a theorem which states that, of all four-dimensional curved spacetime theories of
gravity viable with respect to the four classical weak field tests, only one unique
case, the Y{\i}lmaz theory, has interactive N-body (multiparticle) solutions and this
unique case has no event horizons.  The theorem provides strong theoretical support for
Robertson's explanation of the large energy output of the gamma ray burster GRB 990123
(\cite{5}).  This explanation requires a switch from black holes (a 1-body solution
with horizon) to the case of horizon-free interactive N-body solutions.  In addition to
the good news that the long sought N-body solutions are found, this unique case enjoys
further strong support from other areas of gravitational physics.  This development does
not rule out GRB models with beaming, which can be used if warranted, but it provides a
consistent basis for them, as only in an interactive multiparticle context can such
models be constructed.
\end{abstract}

\keywords{accretion, accretion disks --- black hole physics --- gamma rays: bursts --- methods:
n-body simulations --- stars: neutron --- X-rays: stars}

\section{Introduction}
The difficulty experienced by general relativity in accounting for the prodigious
energy released in the gamma ray bursters, particularly GRB 990123 (\cite{3})
focusses attention on the problems in the treatment of gravitational energy which the
theory has had from its inception. (The quantity proposed by Einstein for the gravitational field 
stress-energy turned out \underline{not} to be a tensor (\cite{8a}).)  The Schwarzschild metric
solution, with the interpretation of "event horizon" and "black hole", limits the mass of a
neutron star to about
$2.8 M_{\sun}$.  But of even greater importance, the existence of interactive N-body solutions
is incompatible with the presence of an event horizon.  N-body interactive solutions, however,
are necessary to describe the
properties of neutron stars, both in the sense of their being a collection of neutrons
(\cite{2a}), and in mergers involving two or more neutron stars as macroscopic N-body systems. 
We also need such solutions to study the formation of neutron stars themselves during
gravitational collapse and the conversion of gravitational energy into radiation in the case of
merging neutron stars or in accretion onto neutron stars in X-ray binaries. These are just some
important examples. All of astrophysics requires interactive N-body solutions which are
not present in general relativity.

A resolution of this problem exists in the relativistic curved spacetime gravitation
theory of Yilmaz which completes the approach initiated by Einstein.  The
fundamental differences between the two theories are presented in this paper
in the form of a theorem about the N-body solutions.  The intent is to provide a brief
theoretical underpinning for Robertson's proposed explanation of GRB 990123 (\cite{5}) and to
note other astrophysical consequences of this new view.

\section{The Problem of Interactive N-Body (Multiparticle) Solutions}
We begin with the obvious remark that, in order to do physics with any set
of objects, we must have more than one object so that we may study their relationships, their
interactions, their scatterings, their coalescence, and so on.  In other words, an acceptable
physical theory must have ``interactive N-body solutions.''  By interactive, we mean the bodies
exert forces on each other, or accelerate in each other's fields when free of constraints.  We
therefore propose to investigate whether a theory of gravity has N-body interactive solutions.

In 1974 the British Canadian mathematician Brian O.J. Tupper has shown that
in any four-dimensional spacetime theory of gravity viable with respect to the four
classical weak field tests the slow motion (sometimes called static) limit $u^i \Rightarrow 0$,
$u_0u^0 \Rightarrow 1$ the field equations are of the form (\cite{7,8})
\begin{equation}
\case{1}{2} G^{\nu}_{\mu} = \tau^{\nu}_{\mu} + \lambda t^{\nu}_{\mu}
\end{equation}
\begin{equation}
\tau^{\nu}_{\mu} = \sigma u_{\mu} u^{\nu},
\end{equation}
\begin{equation}
\sqrt{-g}\sigma = \Sigma_A m_A \delta^3 (\mbox{\boldmath
$x$}-\mbox{\boldmath $x$}_A)
\end{equation}
where $\tau^{\nu}_{\mu}$ is the Einstein ``matter-stress-energy'' tensor,
and
\begin{equation}
t^{\nu}_{\mu} = -\partial_{\mu} \phi \partial^{\nu} \phi +\case{1}{2}
\delta^{\nu}_{\mu} \ 
\partial^{\rho} \phi \partial_{\rho} \phi
\end{equation}
is the Y{\i}lmaz gravitational ``field stress-energy'' tensor and $\lambda$
is an arbitrary
numerical parameter passing through $\lambda = 0$ (Einstein's theory) and
$\lambda = 1$
(Y{\i}lmaz' theory).  The $\phi$ is the low velocity limit of $\phi^0_0$
when $\phi = \ {\rm
trace} \ \phi^{\nu}_{\mu} = \phi^0_0$.  Thus in this limit $\phi$ is a
scalar.

Remarkably, Tupper was able to solve the above equations
\underline{exactly} for arbitrary
$\lambda$ as
\begin{equation}
ds^2 = Adt^2 - A^{-1} (1 - \epsilon^2\phi^2 /4)^2 (dx^2 + dy^2 + dz^2)
\end{equation}
\begin{equation}
\phi = m/r
\end{equation}
where $A = [(1 - \epsilon \phi /2)/(1 + \epsilon \phi /2)]^{2/\epsilon}$,
$\lambda = 1 -
\epsilon^2$.  (We use $2\lambda$ where Tupper used $\lambda$.  Note also
that Tupper used
spherical coordinates whereas we use its transform into cartesian
coordinates in order to
analyze more simply the N-body solutions.)  Tupper noted that for $\lambda
= 0$ (that is,
$\epsilon = \pm 1$) and for $\lambda = 1$ (that is, $\epsilon = 0$) this
metric indeed reduces
to the Schwarzschild and Y{\i}lmaz metrics respectively.  What Tupper did
not emphasize is that,
while the Schwarzschild metric
\begin{equation}
ds^2 = [(1 - \phi /2)/(1 + \phi /2)]^2 dt^2 - (1 + \phi /2)^4 (dx^2 + dy^2
+ dz^2)
\end{equation}
\begin{equation}
\phi = m/r
\end{equation}
is only a 1-body solution $\phi = m/r$, the Y{\i}lmaz metric is an N-body
solution (\cite{9})
\begin{equation}
ds^2 = e^{-2\phi} dt^2 - e^{2\phi} (dx^2 + dy^2 + dz^2)
\end{equation}
\begin{equation}
\phi = \Sigma_A m_A/r_A + C
\end{equation}
where $r_A = |\mbox{\boldmath $x$}-\mbox{\boldmath $x$}_A|$.  It reduces to
$\phi = m/r$ only
as a special case when only one body is present.  Note also that in the
$\epsilon \neq 0$ case
one has an event horizon at $r_{eh} = \epsilon m/2$.  In the Y{\i}lmaz case
$(\epsilon = 0)$
there are no event horizons (no black holes).  Thus we will show that there
is a strict
mathematical anticorrelation between having an event horizon and having
N-body solutions.  

The importance of this result, namely the existence of interactive N-body
solutions in Y{\i}lmaz' theory was not appreciated until recently because it was always assumed (or
hoped) that such N-body solutions would someday be found in general relativity.  However,
despite the many able mathematicians and well motivated groups and alliances working on it during
the last eighty years, no N-body interactive solution has been found in general relativity.
 It is only relatively recently that, using the general $\lambda$-parametric solution
of Tupper, Y{\i}lmaz proved that (Y{\i}lmaz 1987, 1992) they do not exist except when $\lambda =
1$.  Below we present this important theorem and discuss its consequences, including the energy
requirement of GRB 990123.

\section{Proof of the N-Body Theorem}
Since no other generalization than Tupper's $\lambda$-parametric form is viable, and since the
exact solution for arbitrary $\lambda$ is already given, a most interesting thing to do would be
to evaluate (taking Eqs. (1) to (5) into account) both sides of the field
equations with an unspecified $\phi$ and see what happens (what conditions there are in order for
the $\phi$ to be a solution, where the Laplacians occur, etc.).  In fact, something remarkable
happens -- one gets the following \underline{exact} result:

\begin{center}

\begin{tabular}{ccccc}
&&&& \\
&& \it{Left Hand Side} && \it{Right Hand Side} \\
&&&& \\
$\case{1}{2} \sqrt{-g}G^0_0$ & $:$ & $-\Delta \phi + \case{1}{4} (\lambda - 1)
\Omega_{00} + \lambda t_{00}$ &
$=$ & $\Sigma_A m_A \delta^3 (\mbox{\boldmath $x$}-\mbox{\boldmath $x$}_A)
+ \lambda t_{00}$ \\
&&&& \\
$\case{1}{2} \sqrt{-g}G^k_i$ & $:$ & $\case{1}{4} (\lambda - 1) \Omega_{ik} + \lambda
t_{ik}$ & $=$ & $\lambda
t_{ik}$ \\
&&&& \\
\end{tabular}

\end{center}

\noindent where $\Delta$ is the ordinary Laplacian.  $\Omega_{00}$ and
$\Omega_{ik}$ are given
by
\begin{equation}
\Omega_{00} = -2\phi \Delta \phi
\end{equation}
\begin{equation}
\Omega_{ik} = \phi \partial_i \partial_k \phi -3\partial_i \phi \partial_k
\phi +
\delta_{ik} \partial^j \phi \partial_j \phi - \delta_{ik} \phi \Delta \phi
.
\end{equation}
We can see immediately that if $\lambda = 1$ we would have the (needed)
N-body solutions of the
form (computations are done by a Mathematica symbolic manipulation program)
\begin{equation}
\phi = \Sigma_A m_A/r_A + C
\end{equation}
but if $\lambda \neq 1$, $\Omega_{00}$ and $\Omega_{ik}$ will have to
vanish.  The question is,
can we get these two terms to vanish.

Let us first note that, if we ignore the additive constant $(C=0)$, this
can be done in the case
of $\Omega_{00}$.  For, we can take the body as a small sphere of constant
matter density in
which case the potential may be assumed to start from the center as $\sigma
r^2/6$, hence
$\Omega_{00} = -\sigma^2r^2/3$.  Since this expression can be made as small
as one likes, the
$G^0_0$ part of the field equations allows an N-body solution of the type
$\phi =
\Sigma_Am_A/r_A$.  But no matter how hard we try (including the above trick
on its last term) we
cannot get $\Omega_{ik}$ to vanish with an N-body solution where $N$ is
greater than one (for
$N > 1$, $\Omega_{ik}$ has no roots).  But for $N=1$, that is for $\phi =
m/r$, it can easily be
shown that this special case is allowed (because $\Omega_{ik} = 0$ for
$\phi = m/r$) which is the
original $\phi = m/r$ in the Schwarzschild metric.  Thus for the $\lambda =
0$ case we have a
strange situation where the $G^0_0$ component of the field equations allows
an N-body solution
$\phi = \Sigma_Am_A/r_A$ but the $G^k_i$ components of the same equations
do not allow that
solution.  We cannot even argue that, due to nonlinearity, $\Omega_{ik} =
0$ may require a form
different than the $\phi = \Sigma_Am_A/r_A$ because the $\Omega_{00} = 0$ part
already accepts that
form.  This failure is normally overlooked.  The N-body solution is usually
assumed for general
relativity in passing to a Newtonian limit.  But in general relativity the
Newtonian limit is
satisfied only in first order.  Here we are concerned with second order
quantities, $t_{ik}$,
$\Omega_{00}$ and $\Omega_{ik}$.

When we include the requirement of the additive constant the situation gets
worse.  For, in this
case we cannot get even the original Schwarzschild solution. The reason is
that with the
additive constant $C$, neither $\Omega_{00} = -2\sigma C$ nor $\Omega_{ik}
= C\phi_{ik} -
\delta_{ik} \sigma C$ can be made zero independently of $C$ so as to
satisfy the field
equations.  On the other hand, the existence of a $\phi + C$ is of utmost
importance.  The
essence of the additive constant is that if $\phi$ is a solution to the
field equations, then
$\phi + C$ must also be a solution to the same field equations.  This means
that the equations
(and therefore the metric) must not depend on the absolute value of the
potential $\phi$.  They
must depend only on the ``potential differences''.  With the $C$ invariance
also imposed in
general relativity we cannot get any solution at all.  The only way to get
the desired N-body
solutions is to set $\lambda = 1$.  In this unique case the event horizon
disappears and we have
no black holes.  This case allows one to set the zero of the potential
(more generally the
potentials) at the observation point as $\phi (x) \rightarrow \phi (x) -
\phi (x_0)$
where $x_0$ is the position of the observer.  Kinematics then becomes
locally Minkowskian $g_{\mu
\nu} \rightarrow \eta_{\mu \nu}$ which allows one to take over the
measurement procedure of
special relativity locally.  This means that the local vacuum velocity of
light is c
irrespective of accelerations of the observer. 

This interpretation is a
fundamental prediction
of the  $\lambda = 1$ theory and can be tested experimentally (\cite{11}).
To this end, measurements of one-way light propagation times, using 100 picosecond
pulses of laser light and
transported hydrogen maser clocks, have been made over a 20 km East-West
component path.  So far
the results are inconclusive as to whether the East-West and West-East
times are equal or not
on the rotating Earth.  With improved equipment, now available, conclusive measurements can
be obtained
(\cite{1a}).

\section{Interactive Nature of the N-Body Solutions}
As we have emphasized in Section 2, to have N-body solutions is not enough.
 One must also show
that the bodies interact, and in a way consistent with observations.  At
first sight the
linearity of the Poisson equation leading to the N-body solutions may give
the wrong impression
that, in the $\lambda = 1$ case, there may be no interaction between the
bodies hence no
accelerations.  Such a conclusion is not correct, because, in the Newtonian
theory too, we have
such a linear potential and in the Newtonian theory we have interactions
and accelerations. 
Here, as the bodies move, other components $\phi^k_0$, $\phi^k_i$ of the
field also develop.

To exhibit the interactive nature of the N-bodies most clearly we now
introduce the more general
form of the Y{\i}lmaz theory.  Written in a Minkowski background (for
purposes of correspondence
to special relativity, see Note 1) the Y{\i}lmaz theory can be given by two
equations plus a
coordinate (gauge) condition for definiteness, as (\cite{12})
\begin{equation}
\case{1}{2} G^{\nu}_{\mu} = \tau^{\nu}_{\mu} + t^{\nu}_{\mu}
\end{equation}
\begin{equation}
\sigma du_{\mu}/ds = \case{1}{2}  \partial_{\mu}g_{\alpha \beta} (\tau^{\alpha
\beta} + t^{\alpha
\beta})
\end{equation}
\begin{equation}
\partial_{\nu} (\sqrt{-g}g^{\mu \nu}) = 0
\end{equation}
First of all these equations show a relationship between general relativity
and Y{\i}lmaz'
theory.  The former is a truncated case where the $t^{\nu}_{\mu}$ is
removed as in $t^{\nu}_{\mu}
\Rightarrow \lambda t^{\nu}_{\mu}$, $\lambda = 0$.  But the presence of
$t^{\nu}_{\mu}$ is of
crucial importance because there exists an identity which states that
(\cite{Note 2})
\begin{equation}
\case{1}{2} \partial_{\mu} g_{\alpha \beta} (\tau^{\alpha \beta} + t^{\alpha
\beta}) \equiv
(\sqrt{-g})^{-1}  \ \partial_{\nu} (\sqrt{-g}t^{\nu}_{\mu})
\end{equation}
hence the equations of motion can also be written as
\begin{equation}
\sigma du_{\mu}/ds = (\sqrt{-g})^{-1} \ \partial_{\nu}
(\sqrt{-g}t^{\nu}_{\mu}).
\end{equation}
This equation shows clearly the essential point that the $t^{\nu}_{\mu}$ is
the carrier
(mediator) of interactions and in its absence there will be no
accelerations.  This is what we
mean with the requirement of ``interactive N-body" solutions and here we
see that $\lambda = 1$
theory has them.  Multiplying by $\sqrt{-g}$ and integrating over the
volume containing one of
the particles, for example $m_1 = m$,
\begin{equation}
mdu_{\mu}/ds = \Delta \phi \partial_{\mu} \phi = -m\partial_{\mu} \phi
\end{equation}
which is the equation of motion in the slow motion limit.  Since, upon
calculation the term
$1/2 \partial_{\mu} g_{\alpha \beta} (\tau^{\alpha \beta} + t^{\alpha
\beta})$ gives the same
result (note that in this limit $\partial_{\mu} g_{\alpha \beta} t^{\alpha
\beta} = 0$), we
have the geodesic limit $(du_{\mu}/ds = -\partial_{\mu}\phi)$ and the
strong principle of
equivalence satisfied (\cite{12}) 
\begin{equation}
m_i = m_a = m_p
\end{equation}
since $-\Delta \phi = \sqrt{-g}\sigma$ is the density of ``active mass''. 
This calculation also
shows that it will be difficult, if not impossible, to satisfy the ``strong
principle of
equivalence'' without the $t^{\nu}_{\mu}$ because the active mass comes in
by the density
divergence of $t^{\nu}_{\mu}$.  The theory describes interactive
multiparticle dynamics in the
sense of Hamiltonian particle mechanics; the
continuum limit is allowed by statistical
averaging, in which case one needs two or more
functions to describe the details of the
equation of state.

Can there be noninteractive N-body solutions?  It is found that in some
simple
symmetries, extended bodies such as parallel slabs and spherical shells, there may be N of
them even when
$t^{\nu}_{\mu}$ is zero.  However, by the above equations of motion Eq.
(18), the forces between
them, hence also their accelerations are zero (\cite{1}).  If the
$t^{\nu}_{\mu}$ is present,
they do interact (consistent with the Newtonian correspondence).  These
results can be verified
by hand or by computer calculations.

If the solution contains only one object, then, of course, there cannot be
any interaction as
there would be nothing else to interact with.  As to the test-body theories
having a single
central body plus test particles put by hand, they contain an implicit
assumption, namely, the
central body must have infinite inertial mass and finite active mass which
we know is false
and is against the strong principle of equivalence.  Of course, particles
put by hand cannot have
active mass and cannot generate gravitational fields.  Such particles are
called
test-particles.  A test-particle theory violates the universal
interparticle symmetry of
gravitation because the central body is in the solution but the
test-particles are not
(\cite{11a}).

The difference between an N-body theory and a test-body theory shows up
most dramatically in
the calculation of the motions of the planetary perihelia.  Thus for
example perihelion of
Mercury advances 575" per century of which 532" is due to Mercury's
interactions with other
planets and 43" per century to relativistic correction.  The 532"
interactive part is predicted
by the N-body theory but not by the test-body theory since test bodies do
not interact.  The
situation is the same for the other eight planets all of which have even
larger interactive
perihelion shifts.  The $\lambda = 1$ theory predicts the total perihelion
motions in a
seamless way.

     It is usually believed that in papers published in 1938 and 1940 (\cite{3a,3b}) Einstein,
Infeld and Hoffman (EIH) obtained N-body equations of motion in the slow motion limit.  This
belief is unfounded.  As described by P. G. Bergmann in his well-known book (\cite{2b}), the
situation is as follows:  With Eqs. (15.12) on page 230,  Einstein's equations are satisfied in
first order (right hand sides are put to zero in vacuum), but with Eqs. (15.25) on page 234 they
are not satisfied in second order (they are not put to zero in vacuum). They are left
unspecified.  Yet, as stated on page 232,  to obtain the equations of motion one must carry the
field equations to second order.  Thus the question arises: What should these unspecified second
order terms be in order to get the N-body interactive solutions to be used later to obtain the
N-body equations of motion (15.49) on page 240?  It turns out that they cannot be zero, as
Einstein's theory requires.  They rather demand  $\case{1}{2} G^{\nu}_{\mu} =
-t^{\nu}_{\mu}$        in vacuum  where the $t^{\nu}_{\mu}$    is the Yilmaz stress-energy tensor
for the N-body field $\phi =
\Sigma_A m_A/r_A + C $. (The (-) sign is due to the definition of $G^{\nu}_{\mu}$ in
Bergmann's book as the negative of Yilmaz' definition).  In other words, Eqs.(15.49) are true
in Yilmaz' theory and not in Einstein's theory.  In fact, the Yilmaz exponential metric, our
eq.(9), can be derived from the condition that, in the Newtonian limit, the equations of motion
will be of the form (15.49) of Bergmann.

\section{Discussion}
The recent discovery of the gamma ray burster GRB 990123, requiring energies
exceeding the limit allowable by general relativity for neutron star mergers, created an energy
crisis in astrophysics (\cite{3}).  The limiting factor seems to be that, according
to general relativity, a neutron star (or a merger of stars) exceeding a total of 2.8$M_{\sun}$
would become a black hole and thereafter little radiation could escape whereas the energy
required for GRB 990123 seems to be at least 2$M_{\sun}c^2$  to properly account for the gamma
and other emissions.
 In fact, according to the N-body theorem there cannot be such energy producing mergers in
general relativity.  If the obstacle event horizon did not exist, the interaction energy released
from the deeper regions, surfaces, magnetic fields, etc., can provide
the required energy. (Note that the massive neutron stars
can possess magnetic moments -- the "black holes have no hair" theorem does not apply in the
new theory. Note also that radially directed light can always escape, although
substantially redshifted.)  In two recent articles by S. L. Robertson such an explanation is
already proposed (\cite{4,5}).

\underline{Summarizing}: \ a) The long sought N-body interactive solutions
in curved spacetime theory of gravity are found which merits immediate attention in its own
right. \ b) The test-body (1-body) nondynamical metrics are replaced by N-body dynamical
metrics free of event horizons.  A natural explanation of the GRB 990123 energy requirement
becomes possible via a merger of two massive neutron stars (called Y{\i}lmaz stars by Robertson)
which are not black holes.  In the past, in times of great theoretical and observational crises,
like the ones we are now having, patching up old theories did not help.  Instead, a new paradigm
emerged which organized known facts in a more systematic manner as well as overcoming the
prevalent difficulties and predicting new effects.  We may be witnessing here a similar situation
in the equations of general relativity.  In both the field equations and the equations of
motion, the ``matter alone'' paradigm is allowed to go over into a new paradigm ``matter plus
field''.  (More precisely, $\tau_{\mu}^{\nu} \Rightarrow \tau_{\mu}^{\nu} +
t^{\nu}_{\mu}$.)  This change in paradigm makes it possible to treat the GRB 990123 as a merger or
collision of two massive neutron stars, with some beaming if needed, whereas general relativity
seems to be in a bind, since it has only a 1-body solution (a solitary black hole) with which none
of these models is feasible.

Quite independently of the energy crisis at hand this shift in paradigm has
many important consequences in other respects.  a) The theory becomes a standard local
gauge-field theory in curved spacetime. b) It is a dynamical theory (not a test-body theory),
hence the planetary perturbations are treatable in a seamless way along with the relativistic
effects.  c) It does not lead to event horizons, hence physical properties such as magnetic
moments are allowed. d) It has a higher critical mass, hence more energy is available in mergers
and collisions. e) As far as we know, it is quantizable (Y{\i}lmaz 1997, Alley 1995).  These and
other important features will be described in a larger paper in preparation.  \vspace{.25in}
\begin{center}
\it``The hallmark of a successful theory is that it predicts correctly facts
which were not \\known when the theory was presented or, better still, which were then known
incorrectly.''\rm

\hspace{2.2in} Francis Crick  (Life Itself, Simon and Schuster, 1981)
\end{center}

\acknowledgements
We thank Dr. Ching Yun Ren, Mr. Kirk Burrows and Mr. Per Kennet Aschan for
their expert help in computer calculations.  Two of us (HY and YM) would like to thank
Mr. T. Hiruma, President of Hamamatsu Photonics, K.K. for encouragement and dedicate our efforts
to the memory of the late Executive Vice President of Hamamatsu Photonics, Dr. Sakio Suzuki.


\begin{thebibliography}{}
\bibitem[Note 1]{Note 1} Note 1. \ In any Riemannian spacetime theory, in
more general
coordinates than the above, a nontensor $z^{\nu}_{\mu}$, $\tau^{\nu}_{\mu}
\Rightarrow
\tau^{\nu}_{\mu} + z^{\nu}_{\mu}$, $t^{\nu}_{\mu} \Rightarrow t^{\nu}_{\mu}
- z^{\nu}_{\mu}$ may
appear, but this does not harm the coordinate independence of any theory. 
It can be removed
without harm to anything by replacing the ordinary derivative with respect
to the chosen
background by the covariant derivative with respect to the same.  The
reason the Lorentz
background is free of $z^{\nu}_{\mu}$ is that in this case these two
derivatives coincide
(\cite{6,13}).
\bibitem[Note 2]{Note 2} Note 2. \ A corollary of the Bianchi and the Freud
identities
$D_{\nu}(\tau_{\mu}^{\nu} + t^{\nu}_{\mu}) \equiv 0$, $\partial_{\nu}
(\surd -g\tau^{\nu}_{\mu})
\equiv 0$.  If the $t^{\nu}_{\mu}$ is not present $(\lambda = 0)$, both
identities will apply to
$\tau^{\nu}_{\mu}$ and/or $G^{\nu}_{\mu} = 2\tau^{\nu}_{\mu}$, leading to
an overdetermination
(\cite{12}) which is the basic difficulty in the $(\lambda = 0)$ theory. 
The introduction of
the appropriate $t^{\nu}_{\mu}$ removes the difficulty.  Note that Eqs.
(15) and (18) are not
contradictory because when $t^{\nu}_{\mu}$ is absent from the field
equations, the solution
for $\tau^{\nu}_{\mu}$ is such that $\frac{1}{2} \partial_{\mu}
g_{\alpha \beta}
\tau^{\alpha \beta} = 0$.  This is a general result.  A specific exact
solution clarifying this
point has been exhibited (Alley 1994, 1995, Y{\i}lmaz 1994).
\end{thebibliography}
\end{document}